# Self-bias Dependence on Process Parameters in Asymmetric Cylindrical Coaxial Capacitively Coupled Plasma*

J. Upadhyay, [1] Do Im, [1] S. Popović, [1] A.-M. Valente-Feliciano, [2] L. Phillips, [2] and L. Vušković [1]
[1]Department of Physics - Center for Accelerator Science, Old Dominion University, Norfolk, VA 23529, USA
[2]Thomas Jefferson National Accelerator Facility, Newport News, VA 23606, USA

An rf coaxial capacitively coupled $Ar/Cl_2$ plasma is applied to processing the inner wall of superconducting radio frequency cavities. A dc self-bias potential is established across the inner electrode sheath due to the surface area difference between inner and outer electrodes of the coaxial plasma. The self-bias potential measurement is used as an indication of the plasma sheath voltage asymmetry. The understanding of the asymmetry in sheath voltage distribution in coaxial plasma is important for the modification of the inner surfaces of three dimensional objects. The plasma sheath voltages were tailored to process the outer wall by providing an additional dc current to the inner electrode with the help of an external dc power supply. The dc self-bias potential is measured for different diameter electrodes and its variation on process parameters such as gas pressure, rf power and percentage of chlorine in the $Ar/Cl_2$ gas mixture is studied. The dc current needed to overcome the self-bias potential to make it zero is measured for the same process parameters.

We are developing a method to modify the inner surface of superconducting radio frequency (SRF) cavities made of niobium (Nb) by using a coaxial, capacitively coupled rf plasma of argon/chlorine ($Ar/Cl_2$) gas mixture. The surface area asymmetry between inner and outer electrodes due to the coaxial nature creates a negative self-bias potential on the inner electrode plasma sheath. While the self-bias formation in planar geometry can be advantageous for semiconductor wafer processing (as it is placed on the electrode with a self-bias potential), it is detrimental for processing the inner surface of cylindrical structures. The negative dc self-bias potential provides much higher energy to ions bombarding the inner (powered) electrode compared to the ions bombarding outer (grounded) electrode. Since the ion bombardment energy on the outer electrode is very low, it is not feasible to etch the outer electrode without applying a positive dc bias to the inner electrode by an external dc power supply. This power supply drives an additional dc current through the inner electrode in order to bias it positively and change the plasma potential of the bulk plasma, which leads to the increase of ion energy impinging the outer wall.

Particle accelerators and accelerator based light sources in use, or proposed to be built in the future, require SRF cavities. To remove the mechanically damaged layer from the inner surface of these structures, hydrogen fluoride (HF) based chemical methods like buffered chemical polishing (BCP) or electro polishing (EP) are used [1]. Plasma based surface modification will not only reduce the cost and environmental hazards but also open the possibilities of tailoring the surface for better SRF properties.

We previously established the dependence of the etch rate on process parameters for the plasma based method [2] and determined the etching mechanism of Nb [3]. The etch rate non-uniformity due to depletion of radicals along the gas flow direction on process parameters is reported in Ref. [3].

The change in plasma potential and, in turn, the change in ion energy by applying a dc voltage for planar asymmetric rf plasma is reported in Refs. [4-7]. Due to extensive use in the semiconductor industry, planar asymmetric plasma reactors are relatively well understood. The theoretical model for sheath voltage ratio between two electrodes for these discharges and its dependence on their surface area is provided in Refs. [8-10] and the combined operation for dc and rf plasma is described in Ref. [11]. The self-bias dependence on gas pressure and rf power for planar asymmetric plasma is also reported in Ref. [12], however, its behaviour for $Ar/Cl_2$ plasma in coaxial type asymmetry is not known. The quantitative measurement of the self-bias potential for different diameter electrodes and its variation on process parameters is important for any kind of surface modification of the inner surface of three dimensional structures. The additional dc current is needed to bring the negative self-bias potential at the inner electrode to zero or a positive value. This dc current can also be considered as an indication of plasma electron density.

In this paper, we present the dependence of the self-bias potential and external dc current on gas pressure, rf power, and concentration of the chlorine in gas mixture for different diameter inner electrodes. In the following, we describe the experimental setup and discuss the experimental results.

To evaluate the asymmetry in a coaxial type rf plasma reactor, we are using a 7.1 cm diameter and 15 cm long cylindrical tube as an outer electrode, which is part of the vacuum vessel. The inner electrode is of varied diameter (2.5, 3.8 and 5.0 cm) and fixed length of 15 cm. An rf (13.56 MHz) power supply is used to produce the plasma. It is connected to an automatic impedance matching network, which also measures the self-bias potential developed on the inner electrode sheath. The matching network has an additional option to connect a dc power supply in order to vary the dc bias on the inner electrode. The setup includes a dc power supply to provide the current required for each condition to reduce the dc self-bias to 0 V. The gases used were pure argon or chlorine diluted to 15% by adding argon. More details of the experimental setup were described earlier [2, 3].

In this setup, the inner electrode was powered, and the outer electrode was grounded. The ions gain energy in the sheath due to the potential difference between the time averaged potential of bulk plasma $V_p$ and the surface. An increase of negative bias at the inner electrode leaves the plasma potential unchanged, but it can be changed significantly if a positive dc bias is applied [4-7]. The dc coupling allowed a dc current to flow to the powered electrode and to expand the plasma structure to the whole chamber. In the case of low rf power without dc bias, the plasma is confined to the inner electrode, as similarly observed for planar geometry [7]. The etch rate data for different diameter of the inner electrode indicate that the variation in plasma potential was smaller in the case of the smaller diameter electrode and the etch rate of Nb (inner wall of outer electrode) was reduced.

The negative self-bias potential developed across the inner electrode sheath for all three inner electrode diameters was measured at different gas pressure, rf power and two gas compositions. The required dc current to bring this potential to zero or positively biased at a certain value was also recorded.

The self-bias dependence on pressure was measured at rf power of 25, 50, 100 and 200 W. The variation of self-bias potential with the pressure for Ar plasma using the three diameter electrodes is shown in Fig. 1 at rf power of 100 W. The trends of the curves for the other measured rf powers are similar. The errors in measurements of $Cl_2$ concentration was 2%, in rf power 3 W, in pressure 4 mTorr, in d.c. bias 2 V and in dc current 5 mA.

The self-bias displayed in Fig. 1 shows two distinct pressure regimes, one below 150 mTorr and other above 150 mTorr. Though the self-bias potential is negative, it is plotted on the positive axis for convenience. The increase of the self-bias at low pressures could be explained by the expansion of the plasma volume to the grounded area [12]. The rf power dependence of self-bias for different diameter electrodes is shown in Fig. 2.

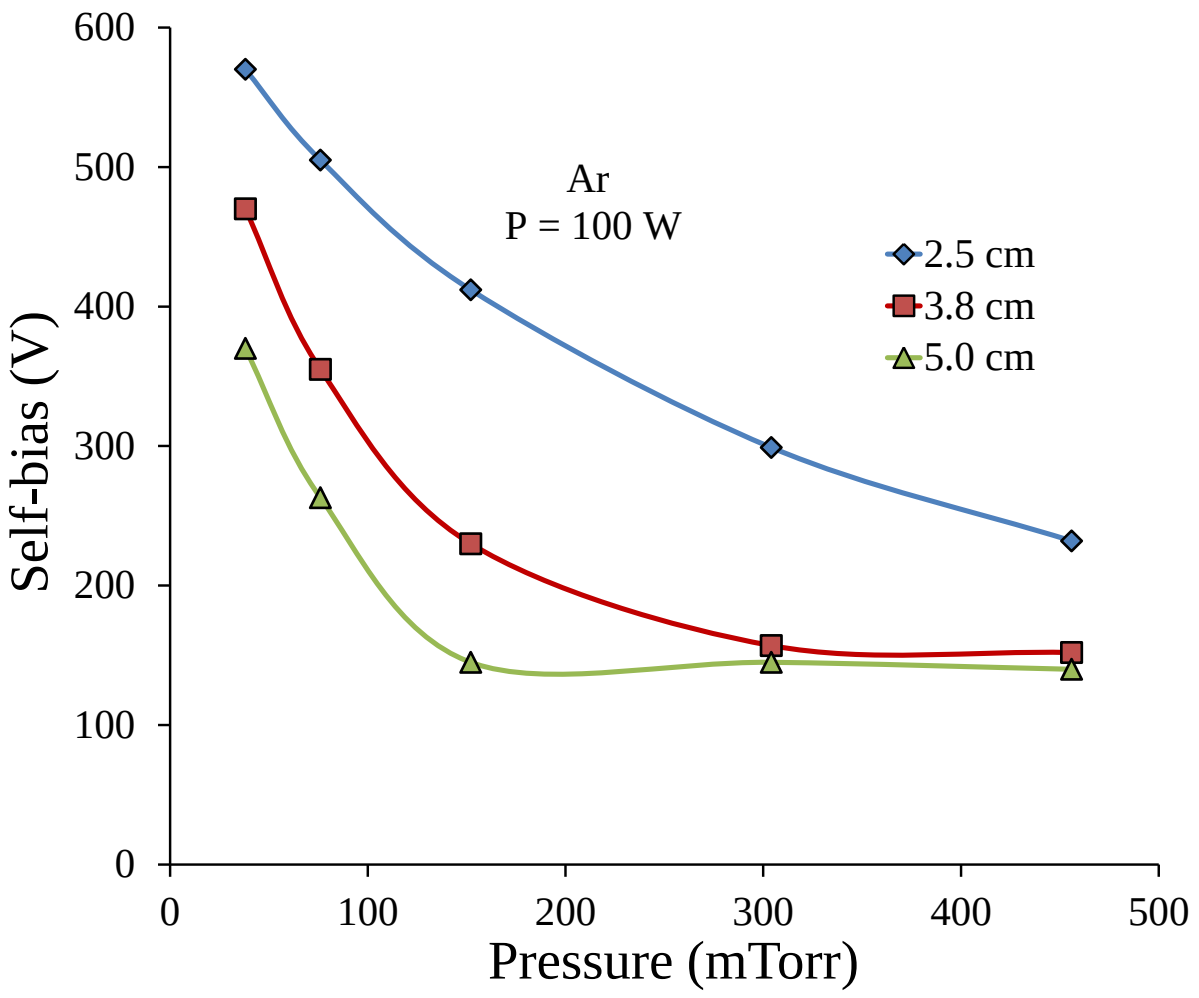

FIG. 1. Self-bias dependence in the Ar plasma on the pressure for different diameter electrodes. Solid lines are visual guidelines.

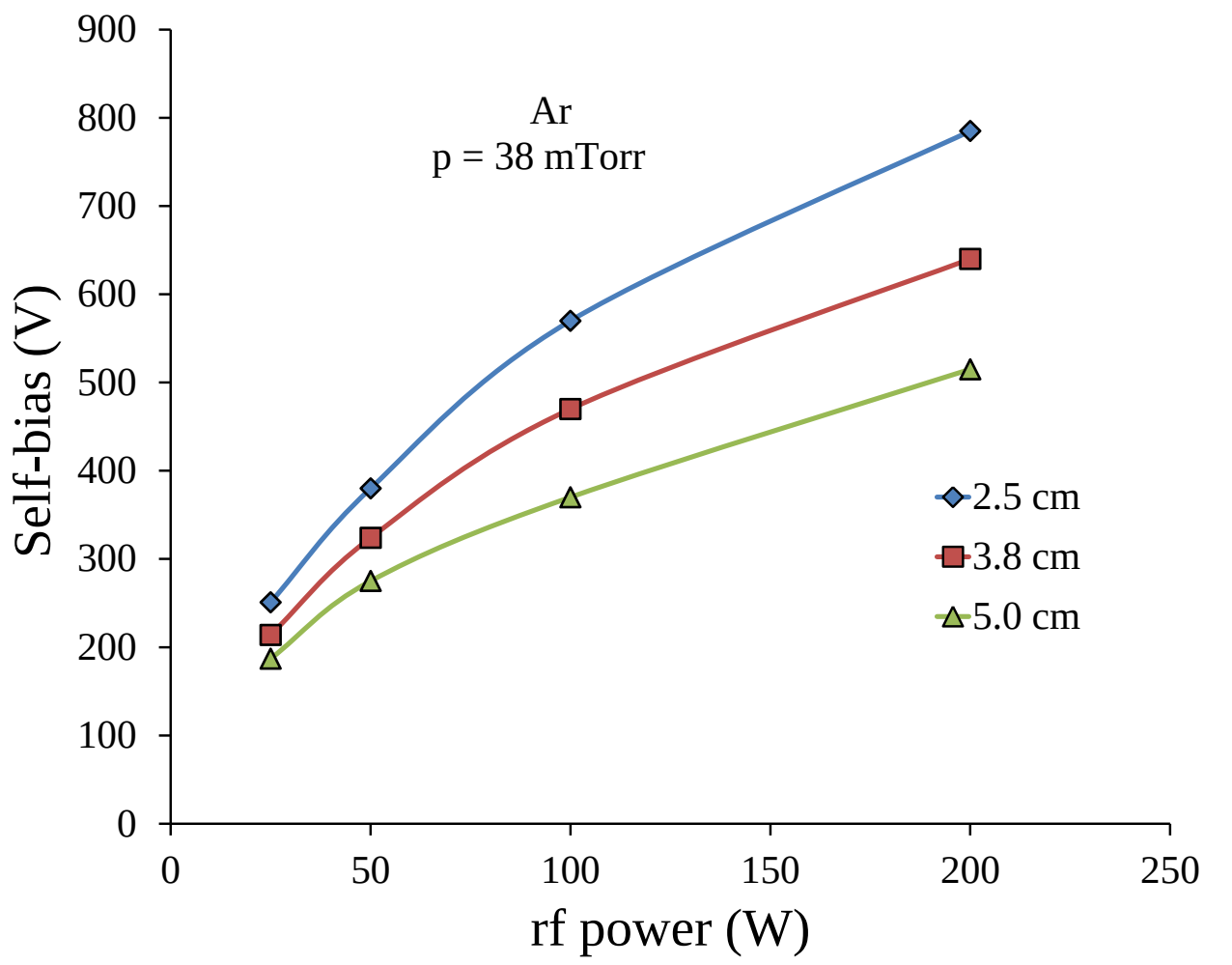

FIG. 2. Self-bias dependence in the Ar plasma on the rf power for different diameter electrodes. Solid lines are visual guidelines.

The increase in the self-bias with the rf power is not only due to the expansion of the plasma but also the rf voltage increase. The fit to these curves shows almost square root dependence on the rf power indicating that all the rf voltage is dropping on the inner electrode sheath as a dc bias [13].

The self-bias voltage varies for pure Argon and $Ar/Cl_2$ mixture. Figure 3 shows the variation of self-bias voltage with rf power for the inner electrode diameters of 2.5 and 3.8 cm. The self-bias voltage for $Ar/Cl_2$ gas mixture is smaller than for pure Ar for lower rf power but larger at higher rf power. This behavior could be partially explained with the relative electron density decrease at lower power in $Ar/Cl_2$ plasma compared to the positive ion density due to large plasma electronegativity. The electron density is almost equal to the positive ion density at higher rf power density as reported in Ref. [14]. This property of chlorine plasma is reflected in the self-bias voltage variation at higher rf power.

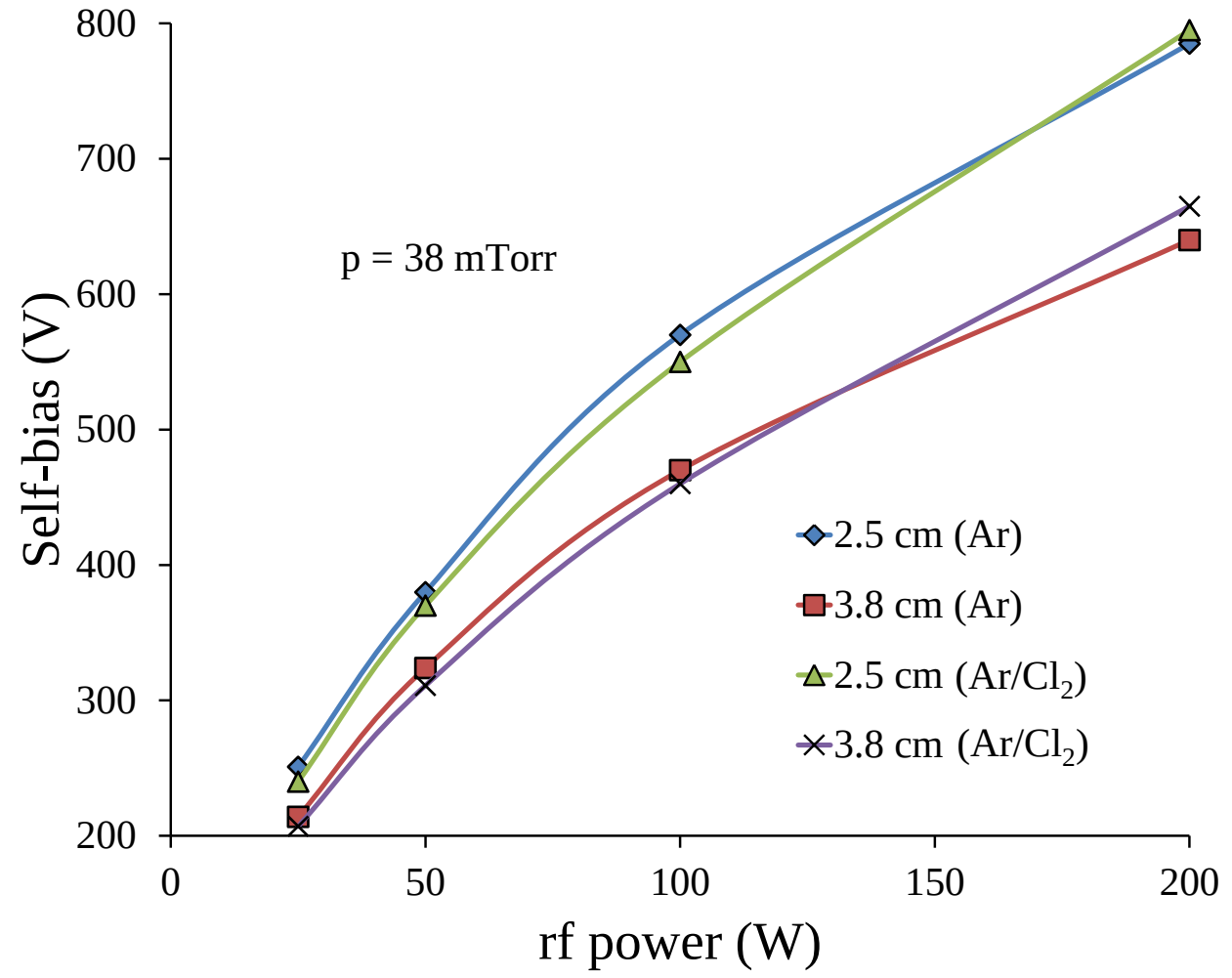

FIG. 3. Self-bias dependence on the rf power for different diameter electrodes for Ar and $Ar/Cl_2$ plasma. Solid lines are visual guidelines.

In the case of the inner electrode diameter of 5.0 cm, the plasma volume is smaller and equal density of electrons and positive ions is reached earlier as reported in Ref. [14]. However, the electron temperature is higher in the $Ar/Cl_2$ plasma compare to the Ar plasma, consequently the self-bias potential in the $Ar/Cl_2$ plasma is higher than the pure Ar plasma, which is shown in Fig. 4.

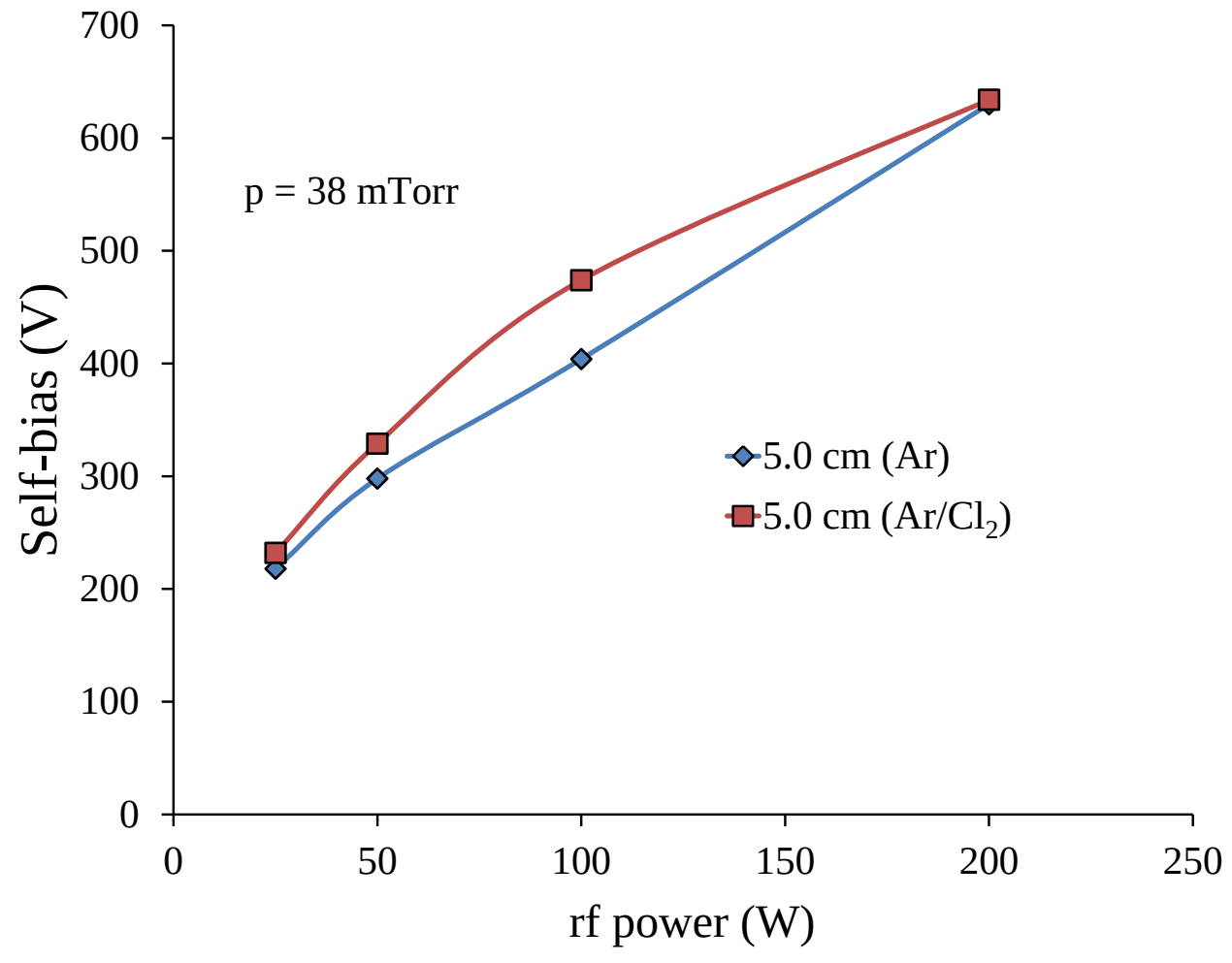

FIG. 4. Self-bias dependence on the rf power for electrode diameter of 5.0 cm for pure Ar and $Ar/Cl_2$ mixture. Solid lines are visual guidelines.

The variation of the self-bias with pressure at fixed rf power for pure Ar and $Ar/Cl_2$ plasma for inner electrode diameter of 5.0 cm is shown in Fig. 5.

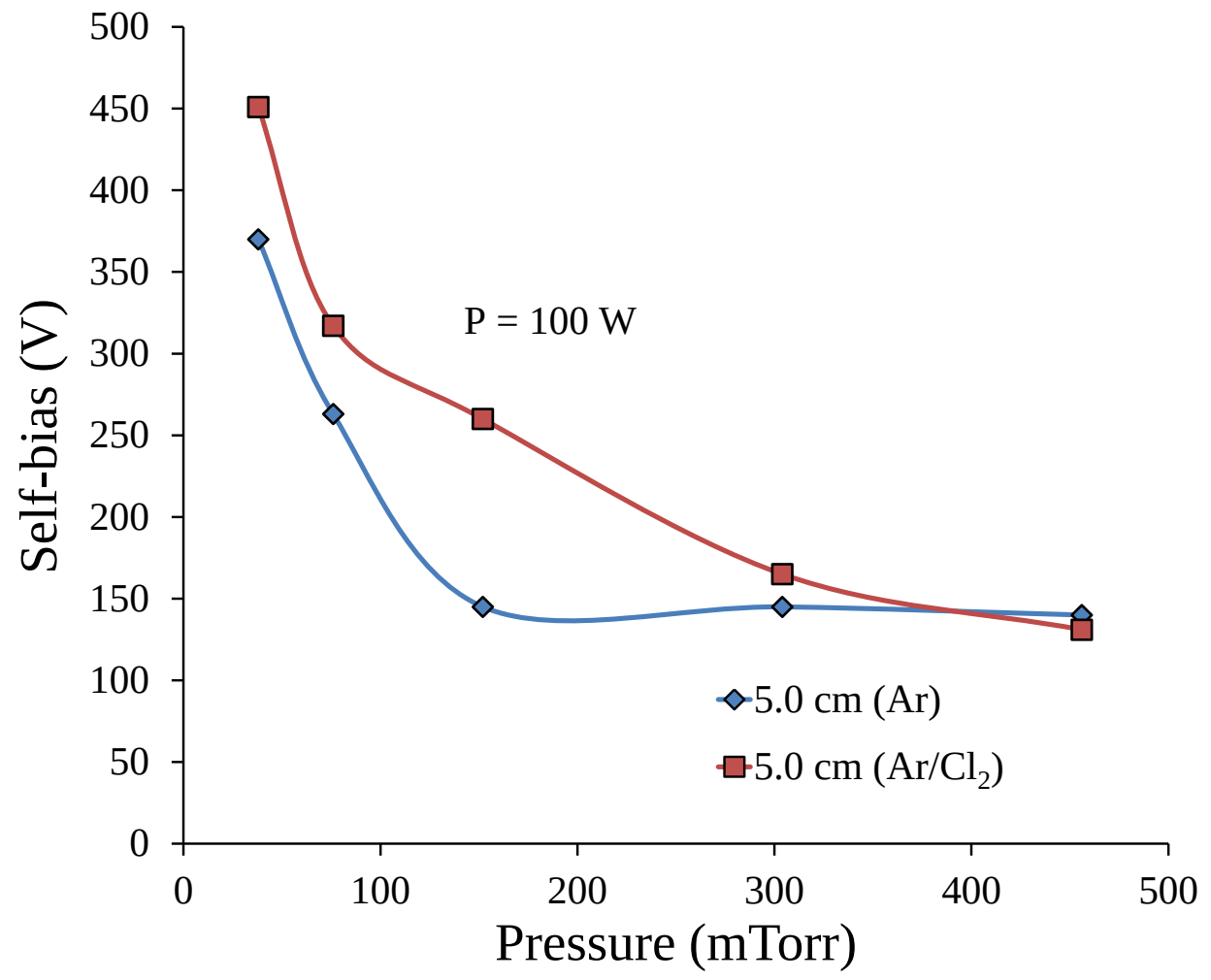

FIG. 5. Self-bias dependence on the pressure for pure Ar and $Ar/Cl_2$ mixture plasma for the same diameter electrode. Solid lines are visual guidelines.

Figure 5 shows that, due to lower electron density and higher electron temperature in $Ar/Cl_2$ plasma the self-bias voltage is higher compared to pure Ar plasma. This higher self-bias voltage shows that the addition of electronegative gases such as $Cl_2$ increases the asymmetry in rf plasma reactors.

Additional dc current has to be provided with the help of an external power supply to bring the dc bias to zero or positive value in order to increase the plasma potential for all three inner electrodes. The current, needed to increase the negative dc self-bias to 0 V, could be treated as an indicator of electron density in the plasma. The dc current required to lift up the self-bias voltage to zero for pure Ar and $Ar/Cl_2$ plasmas, is plotted in Fig. 6.

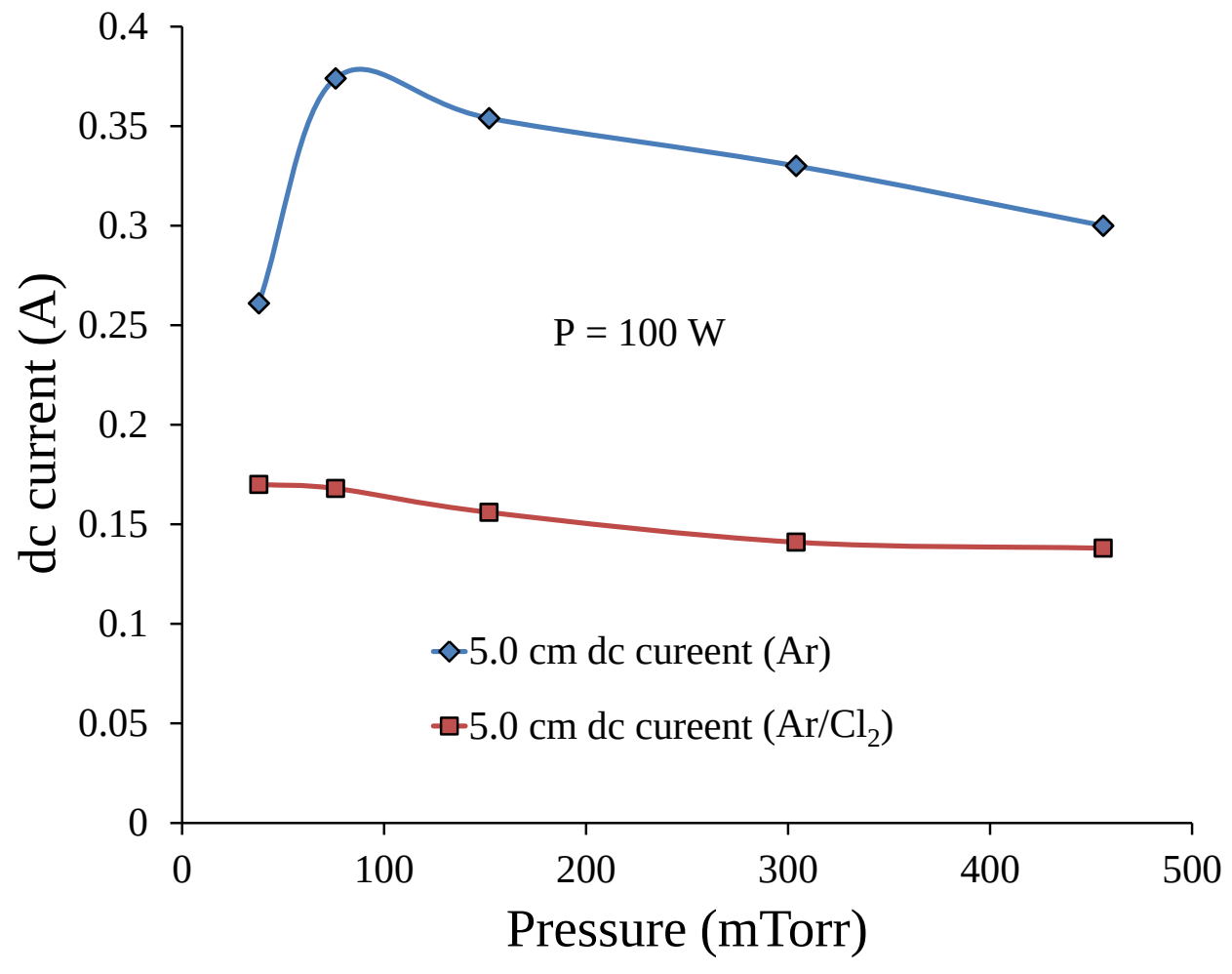

FIG. 6. The dc current variation on the pressure for 5.0 cm diameter electrode for Ar and $Ar/Cl_2$ plasma. Solid lines are visual guidelines.

Figure 6 shows that much less current has to be provided in the case of $Ar/Cl_2$ plasma, as the electron density is approximately an order of magnitude lower in $Cl_2$ plasmas compared to Ar plasma, which was reported earlier in [13]. The similar trend is recorded for the other two diameter electrodes for all the pressure and power conditions. The addition of $Cl_2$ decreases the electron density in the plasma. The increase in dc bias voltage increases the current provided by the dc power supply. The dc bias voltage variation with dc current is shown in Fig. 7.

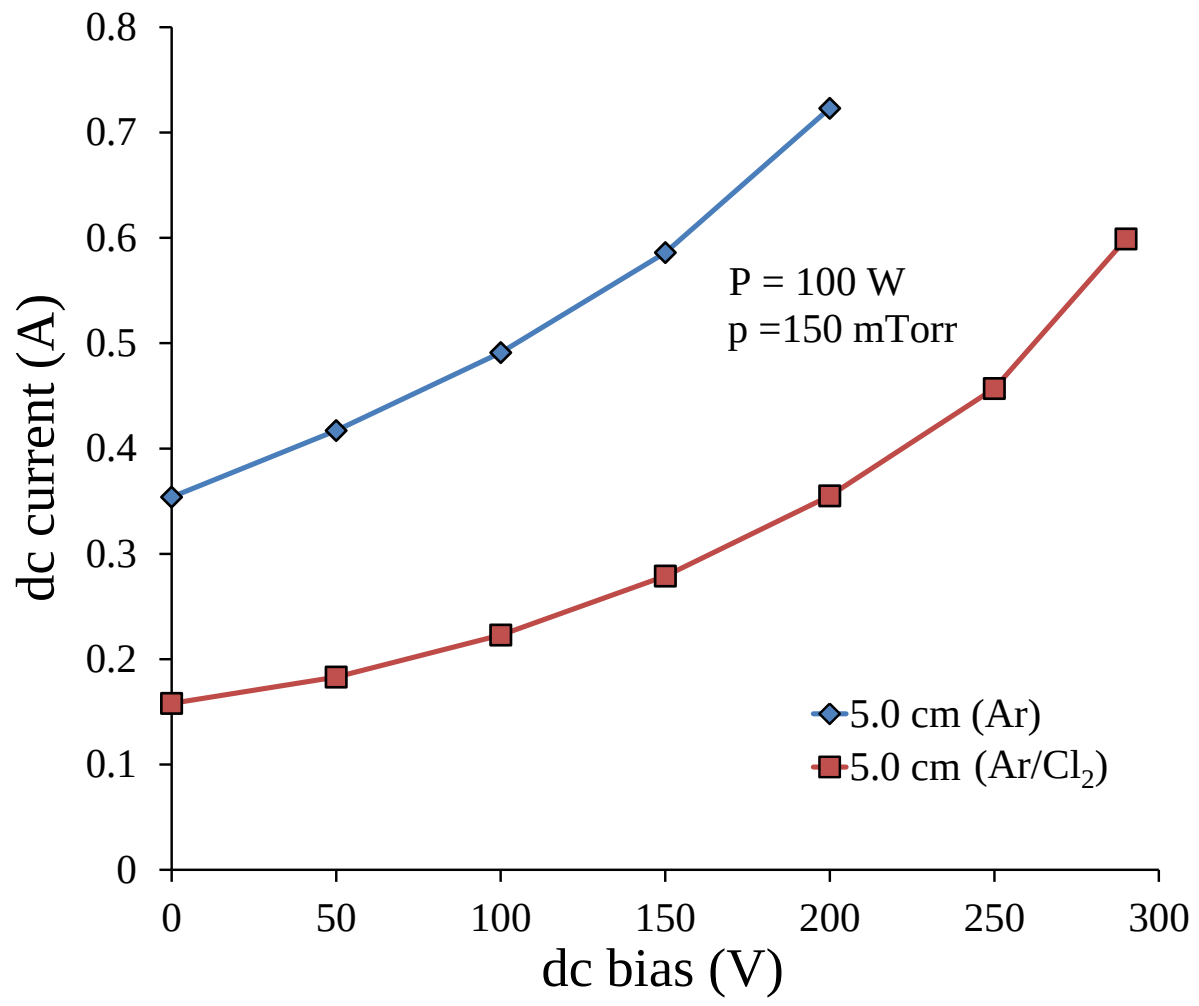

FIG. 7. The dc current variation on the dc bias voltage at fixed pressure and fixed rf power for Ar and $Ar/Cl_2$ plasma. Solid lines are visual guidelines.

The increase in dc current with biased voltage indicates that positive dc biasing not only increases the plasma potential but also the plasma density, as also shown in case of planar geometry [12].

This study presents the self-bias dependence in cylindrical coaxial capacitively coupled plasma on gas pressure and rf power using different diameters of the inner electrode. The self-bias data show a difference between Ar and $Ar/Cl_2$ plasma and an explanation for this difference is offered. This study also presents the variation in dc current required for bringing the self-bias potential to zero or positive for Ar and $Ar/Cl_2$ plasma and the role of plasma density in this variation. Results also leads to the observation that positive dc bias increases the plasma density together with plasma potential, which is beneficial in etching the inner surface of the outer cylindrical wall of a cavity.

*This work is supported by the Office of High Energy Physics, Office of Science, Department of Energy under Grant No. DE-SC0007879. Thomas Jefferson National Accelerator Facility, Accelerator Division supports J. Upadhyay through fellowship under JSA/DOE Contract No. DE-AC05-06OR23177.